\documentclass[twocolumn,showpacs,preprintnumbers,amsmath,amssymb]{revtex4}

\newcommand{\be}{\begin{equation}}
\newcommand{\ee}{\end{equation}}

\newcommand{\bc}{\begin{center}}
\newcommand{\ec}{\end{center}}
\newcommand{\bdm}{\begin{displaymath}}
\newcommand{\edm}{\end{displaymath}}

\newcommand{\forget}[1]{}

\newcommand{\bea}{\begin{eqnarray}}
\newcommand{\eea}{\end{eqnarray}}

\usepackage{graphicx}
\usepackage{dcolumn}
\usepackage{bm}

\begin{document}

\preprint{APS/123-QED}

\title{Front pinning in capillary filling of chemically coated channels}

\author{F. Diotallevi, S. Chibbaro, S. Succi,}
 \affiliation{ Istituto per le Applicazioni del Calcolo CNR 
    V. Policlinico 137, 00161 Roma, Italy.}
 \email{f.diotallevi@iac.cnr.it}

\author{ A. Puglisi}
 \affiliation{CNISM and Dipartimento di Fisica, Universita' La Sapienza, P.le A. Moro 2, I-00185 Rome, Italy.
}%

\author{L. Biferale}
 \affiliation{ Dipartimento di Fisica e INFN, Universita' di Tor Vergata, Via della Ricerca
    Scientifica 1, 00133 Rome, Italy
}%

\date{\today}

\begin{abstract}
The dynamics of capillary filling in the presence
    of chemically coated heterogeneous boundaries is investigated,
    both theoretically and numerically. In particular, by mapping the equations of
    front motion onto the dynamics of a dissipative driven oscillator,
    an analytical criterion for front pinning is derived, under the condition of diluteness of the coating spots. The
    criterion is tested against two dimensional Lattice Boltzmann
   simulations, and found to provide satisfactory agreement
   as long as the width of the front interface remains much thinner than the typical heterogeneity scale of the chemical coating.
\end{abstract}

\maketitle

\section{
\label{intro}
Introduction}
The physics of capillary filling, originated with the
pioneering works of Washburn \cite{washburn} and Lucas \cite{Lucas},
has provided a constant source of interesting
problems in fluid dynamics~\cite{degennes,dussain}. Recently, with the burgeoning growth
of theoretical, experimental and numerical works on micro- and
nano-fluidics, the problem attracted a renewed interest
\cite{washburn_rec,tas}.  Capillary filling is a
typical ``contact line'' problem, in which  the subtle non-hydrodynamic
effects taking place at the contact point between liquid-gas and solid
phase allow the interface to move, pulled by capillary forces and
contrasted by viscous forces. Usually, only the late asymptotic stage
is studied, leading to the well-known Lucas-Washburn law \cite{washburn},
which predicts the following relation for the position $z(t)$ of the
moving interface inside the capillary:
\begin{equation}
\label{wash_base}
z^2(t) - z^2(0)= \frac{\gamma H cos(\theta)}{3 \mu }  t 
\end{equation}
where $\gamma$ is the surface tension between liquid and gas, $\theta$
is the {\it static} contact angle, $\mu$ is the liquid dynamic viscosity, $H$
is the channel height and the factor $3$ depends on the geometry of
the channel. Here, we focus on a two dimensional geometry given by two infinite
parallel plates, separated by a distance $H$ (see fig. \ref{coat}A). 

One of the practical problems associated with the Lucas-Washburn law
is the monotonically vanishing speed (Eq.~(\ref{wash_base}) implies
$\dot{z} \simeq t^{-1/2}$) of the liquid front in the late stage of
the process.  The inertial effects that characterize the initial stage
of the filling process, decay exponentially in time, thus leading to a
higher sensitivity of the front motion to chemical or geometrical
defects deposited along the walls. As a result, the front
  often pins, since even the smallest hydrophobic spot represents a
  potential barrier that cannot be overcome by a front moving at
  a vanishingly small speed. 

In this work we deal with ``chemical'' coating, modelled through a
position dependent contact angle, $\theta(z)$. In particular, we focus
on a dilute regime, where defects (hydrophobic regions) are sparsely
deposited on the walls, with a typical inter-distance larger than the
length scale associated with inertial effects in the flow. This
approach is interesting from the practical standpoint, because, in
actual experiments, hydrophobic coating minimizes impurities
adsorption at the wall. Therefore, it is of importance to provide a
quantitative relation between the average fraction and chemical
properties of the hydrophobic surface, and the maximum length reachable
by the interface along the capillary. Indeed, this immediately
reflects into minimal requirements for the control of the coating
process, or for the external supply needed to support front
propagation up to a given distance.

\section{
\label{setup}
Geometrical Setup}
\begin{figure}[htb]
\centering\scalebox{0.6}{\includegraphics[clip=true]{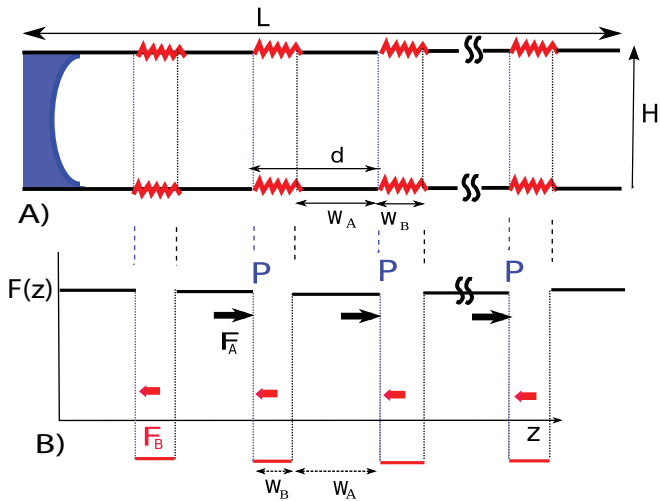}}
\caption{Cartoon of the heterogeneous coating.  Top: both the walls of
  the channel are partitioned in an alternate sequence of strips A and
  B, with different chemical properties ($\theta_A$ and $\theta_B$)
  and different lengths ($w_A$ and $w_B$). Bottom: force $f(z)$ as a
  function of $z$. In the hydrophilic regions ($A$) the front is
  pushed rightwards, while in the hydrophobic ones ($B$) it is pulled
  leftwards.}
\label{coat}
\end{figure}

We consider a fluid
 penetrating a
channel of length $L$ with periodic hydrophobic
 spots, as depicted
in Fig.~\ref{coat}A. As we shall see in the following, the periodicity is not
an important restriction, as long as the distance between two
consecutive hydrophobic spots is large enough, i.e. if the
"diluteness" requirement is fulfilled. 
 The channel is coated with
an alternate sequence of strips A and B, of
 length $w_A$ and $w_B$
respectively, uniformly repeated along the
 channel length. Regions A
and B have different wetting properties,
 characterized by the
contact angles $\theta_A$ and $\theta_B$. In the
 following, we
consider the case $\theta_A<\pi/2$ and
 $\theta_B>\pi/2$, a situation
corresponding to a periodic sequence
 of attracting (hydrophylic) and
repelling (hydrophobic) sites
 (Fig.~\ref{coat}B).
 
\par In this Letter, we investigate the dependence of the front speed
on the two lengths $w_A, w_B$ and on the wetting angles $\theta_A$,
$\theta_B$.  In particular, we show that it is possible to reformulate the filling problem in terms of a
damped-forced oscillator, and we develop a procedure that allows for an analytical
prediction of the pinning position.  This analytical
 criterion is
tested against direct numerical simulations based
 on the Lattice
Boltzmann equation.

\section{
\label{sec:level3}
The Lucas-Washburn Equation}
The Lucas-Washburn equation describing a liquid of density $\rho_l$ and
dynamic viscosity $\mu_l$, filling a capillary channel up to a distance
$z$, reads as follows:
\begin{equation}
 z \frac{d^2 z}{d t^2} + \left (\frac{d z}{d t}\right )^2 = -\eta z \frac{d z}{d t}+f(z).
\label{washgen}
\end{equation}
The LHS represents the inertial terms, while the RHS reports the
viscous and capillary forces. The parameter $\eta=\frac{12 \mu_l}{\rho_l  H^2}$ in front of the viscous term defines a characteristic time, $ t_\eta =
\eta^{-1}$, beyond which the dynamics is controlled by the balance between the viscous damping and the
capillary force $f(z)= \frac{2 cos(\theta(z)) \gamma}{\rho_l H}$.
Here $H$ denotes the height of the
channel, supposed for simplicity 2d. 

Eq.~(\ref{washgen}) is exact only in the limit of liquid
 filling an
empty capillary. In the case of a liquid-gas system, the
 density
ratio of the two phases $\alpha=\frac{\rho_g}{\rho_l}$ must be
 taken
into account \cite{condmat1,condmat2,napoli}.
 
 As we shall see
shortly, Eq.~(\ref{washgen}) stems from the balance
 between the
total momentum change inside the capillary and the force
 acting on
the liquid system.  The explicit expression of the viscous
 drag is
obtained assuming a developed Poiseuille profile inside the
 flow,
with speed $u(y) = 6 \frac{ \bar{u} }{H^2} y(H-y)$.  In
 addition,
the front speed is identified with the average flow speed in
 the
channel, namely $\dot z \equiv \bar{u} =1/H \int_0^H u(y) dy$.
 
For the case depicted in Fig.~\ref{coat}, the forcing $f(z)$ in the
RHS of Eq.~(\ref{washgen}) is a piecewise constant function,
alternating between positive and negative values, as
 shown in
Fig.~\ref{coat}B.
 Upon introducing the new variable $q=z^2/2$,
Eq.~(\ref{washgen}) simplifies to:
\begin{equation}
\label{basic}
\frac{d^2 q}{d t^2} = -\eta \frac{d q}{d t} +F(q)
\end{equation}
where, due to the non-linear change of coordinates, $F(q)$ is now
a {\it non-periodic} piecewise constant function of $q(z)$. Note that $F(q)$
has the dimensions of a squared velocity, more precisely $F(q)=2 V_{cap}V_d \cos(\theta(q))$, where $V_{cap}=\gamma/\mu_l$ is the capillary speed, and $V_d=\nu_l/H$ is the diffusive speed, being $\nu_l=\mu_l/\rho_l$ the kinematic viscosity of the liquid.

Eq.~(\ref{basic}) describes the motion of a forced damped oscillator
with total energy $E(q)=K(q)+V(q)$, where $K(q)=\frac{1}{2}\left
  (\frac{dq}{dt}\right)^2$ and $V(q)$, defined by the expression $-\frac{dV}{dq}=F(q)$, are the
``kinetic'' and ``potential'' energies, respectively. In the
particular case $F(q)=const$ (homogeneous, non-coated channel), integrating Eq.~(\ref{basic}) once, leads t:
\be
\label{solbasic}
\dot{q}(t)=F t_{\eta}+\left ( \dot{q}(t_0)-F t_{\eta} \right ) \exp \left (-\frac{(t-t_0)}{t_\eta}\right ).
\ee
which, in the limit $t-t_0 \gg  t_\eta$, tends to the asymptotic value
\be
\label{vinf}
\dot{q}(\infty)=v_{\infty}=F t_{\eta} = \frac{\gamma H cos(\theta)}{6 \mu_l}.  
\ee 
Taking into account the definition $q=z^2/2$, it is readily checked
that Eq.~(\ref{vinf}) is equivalent to Eq.~(\ref{wash_base}).

\section{Chemical Coating}
Let us now focus the  attention on the case of a dilute dispersion of
hydrophobic spots along the channel walls. In other words, we shall
always assume that the length of all hydrophlic spots, $w_A$ is much
larger
than that of the hydrophobic ones, $w_B$. Moreover, we also assume the natural
condition that the length of hydrophilic spots is large enough to
assure that inside each pulling region the front reaches its
asymptotic Washburn-Lucas velocity, $v_{\infty}$, given  by expression (\ref{vinf}). The  critical
minimal lenght, $w_A^{crit}$, 
 of the hydrophilic region which satisfies the above
requirement is easily estimated out of Eq.~(\ref{vinf}). The requirement is that in a time $\Delta t \sim {\cal O}(t_{\eta})$ the front has travelled at
least a  distance $\Delta q = (w_A^{crit})^2$, leading to:
\be
\label{crit_hydro2}
w_A^{crit} \sim  t_\eta \sqrt{F_A}.
\ee
Typical values for water-air experiments are $\mu=10^{-3}$ kg
m$^{-1}$s$^{-1}$,$ \gamma=72 \;10^{-3}$ kg/s$^2$, yielding a
minimum sparseness
requirement of the order of  $w_A^{crit}  \sim 10 \mu m$ for $H=10\mu
m $.
Of course,if the hydrophilic pulling regions are not long
enough, the front dynamics is always in the inertial-transient 
region of the dynamical evolution of Eq.~(\ref{basic}), and the overall
dynamics becomes less universal and strongly dependent on the coating details.

\section{Pinning Criterion}
\label{sec:pinning}
\begin{figure}[htb]
\centering\scalebox{0.75}{\includegraphics[clip=true]{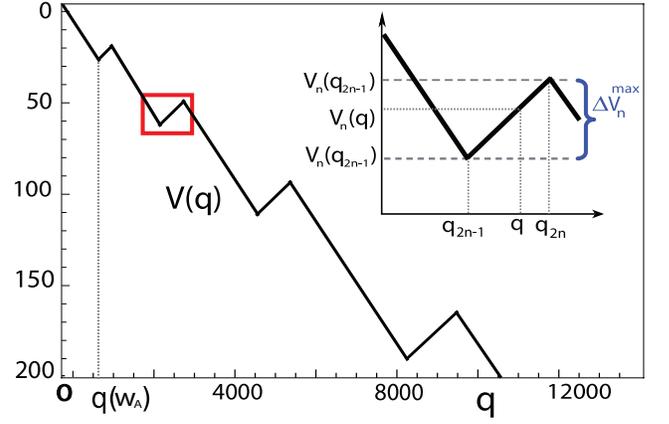}}
\caption{Potential $V(q)$ as a function of $q$: the ratchet form is clearly visible, as well as the position-increasing potential barriers.}
\label{potential}
\end{figure}

Given the definition of $q(z)$, we note that both
the potential barrier to be overcome in
order to "jump" over the hydrophobic obstacle, as well as the extension of
the regions in which the front is accelerated
(Fig.~\ref{potential}), are increasing functions of $z$ (and
consequently of $q$). Therefore, if the front manages to approach its
asymptotic velocity before encountering the first barrier, it will
attain the same asymptotic velocity at the bottom of each of the
subsequent potential wells.

This observation opens the way to a simple prediction of the number of
barriers the front is able to overcome before being pinned. Given that the
kinetic energy $K^A_{\infty}=\frac{(v^A_{\infty})^2}{2}$ accumulated
by the particle as it reaches the bottom of the wells, remains
constant for all the $n$ jumps, the final pinning position is
fixed by the condition $K^A_{\infty}<\Delta V_n^{max}$, where
$\Delta V_n^{max}$ is the ever-increasing heigth of the $n$ potential
wells.

Let us focus on the dynamics during the climbing of the $n$-th
hydrophobic barrier, located between positions $q_{2n-1}$ and $q_{2n}$
(inset of Fig.~\ref{potential}) and denoted by $\Delta
V_n(q)=V(q_{2n})-V(q)$. While the value of the potential barrier
decreases from $\Delta V_n^{max}= V(q_{2n})- V(q_{2n-1})$ to $\Delta
V(q_{2n})=0$, kinetic energy is partly converted into potential energy and partly
dissipated by viscous forces.

If the coordinate $q$ lies in the range $q_{2n-1}< q <q_{2n}$, the temporal evolution of potential and kinetic energy obeys the following equations:
\be
\label{eq1}
\frac{d}{d t}\Delta V_n(q)  = F_B \sqrt{2 K(q)}.
\ee
and
\be
\label{eq2}
\frac{d}{d t}K(q) = -2 \eta K(q) +F_B \sqrt{2 K(q)}  .
\ee 

Equation~(\ref{eq2}) is obtained from Eq.~(\ref{basic}), by multiplying
by $\dot{q}=\sqrt{2 K(q)}$.  Dividing Eq.~(\ref{eq1}) by Eq.~(\ref{eq2}) and
integrating, we obtain \be \int_{K^A_{\infty}}^K \frac{d
  K(q)}{1-\frac{\sqrt{2 K(q)}}{v^B_{\infty}}}= \int_{\Delta
  V_n^{max}}^{\Delta V_n} d\Delta V_n(q) \ee where
$v^B_{\infty}=\frac{\gamma H cos(\theta_B)}{6 \mu_l}$.  This equation, expressing the
kinetic energy $K$ as a function of the potential barrier $\Delta
V_n(q)$, is easily solved, to yield 
\bea
&\Delta V_n - \Delta V_n^{max} =\\
& v^B_{\infty} \left( (v^A_{\infty}-\sqrt{2 K})+v^B_{\infty}
  \log(\frac{\sqrt{2 K^A_{\infty}}-v^B_{\infty}}{\sqrt{2
      K}-v^B_{\infty}}) \right),
 \eea
where by definition $v^A_{\infty}=\sqrt{2 K^A_{\infty}}$.
 The front is pinned whenever two conditions are simultaneously
met: kinetic energy is depleted, $K=0$, and the front position lies
inside the hydrophobic region, i.e. $\Delta V_n>0$.  

This yields 
\be
\Delta V_n= \Delta V_n^{max}+v^B_{\infty} (v^A_{\infty}+v^B_{\infty} \log \left (1-
\frac{\cos(\theta_A)}{\cos(\theta_B)})\right )>0,
\label{sol1}
\ee  
where  $\Delta V_n^{max}=-F_B(q_{2n}-q_{2n-1})$ is the potential barrier 
of the $n$-th hydrophobic spot. Expressing $\Delta V_n^{max}$ as an explicit function of $n$ one obtains:
\begin{equation}
\label{eq:deltamax}
\Delta V_n^{max}=-F_B w_B \left [n (w_A+w_B) - \frac{w_B}{2} \right ] \simeq -F_B w_B L_p
\end{equation}
where $L_p= n(w_A+w_B) - w_B$ is the pinning length. Inserting Eq.(\ref{eq:deltamax}) into Eq.(\ref{sol1}),  after
some algebra, the following prediction for the dimensionless pinning length $\tilde{L}_p =L_p/H$ is obtained:
\begin{equation}
\label{eq:final}
\tilde{L}_p\simeq \frac{C_{cap}\;\cos(\theta_A)}{\tilde{w}_B}\left[ 1-  \frac{\log \left(1+
\frac{\cos(\theta_A)}{|\cos(\theta_B)|}\right)}{\frac{cos(\theta_A)}{|cos(\theta_B)|}} \right].
\end{equation} 
Here $C_{cap}=V_{cap}/(72 V_d)$ is a constant depending on
the capillary parameters, $\tilde{w}_B=w_B/H$ is the dimensionless
hydrophobic length, and $-1<\cos(\theta_B)<0$.  Eq.~(\ref{eq:final})
uniquely identifies the pinning region as a function of the coating
properties and of the typical extension of the hydrophobic spots,
$\tilde{w}_B$.  As intuition suggests, the pinning length
$\tilde{L}_p$ decreases for intense hydrophobic coating, i.e. for
large hydrophobic strength ($\cos(\theta_B)\rightarrow -1$) and for
high fraction of geometrical space covered by the hydrophobic
component (large $\tilde{w}_B$).  Eq.~(\ref{eq:final}) highlights two
distinguished limits, that is {\it fast-filling}
($\frac{\cos(\theta_A)}{|\cos(\theta_B)|}>>0$) and {\it slow-filling}
($\frac{\cos(\theta_A)}{|\cos(\theta_B)|}\rightarrow 0$). In the former
case Eq.~(\ref{eq:final}) delivers $\tilde{L}_p\simeq
C_{cap}\;\cos(\theta_A)/\tilde{w}_B$, while in the latter the pinning
length vanishes like $\tilde{L}_p\simeq C_{cap}\;\cos(\theta_A)^2/(2
\tilde{w}_B\;|\cos(\theta_B)| )$. This asymptotic expressions show that
in the fast filling regime the pinning length $\tilde{L}_p$ grows
linearly with the hydrophilic $\cos(\theta_A)$, independently of the
hydrophobic strength, and stays finite for any non-zero value of $\tilde{w}_B$. On the other hand, in the slow-filling regime,
the pinning length vanishes with the square of the hydrophilic
$\cos(\theta_A)$ and is inversely proportional to the hydrophobic
one $\cos(\theta_B)$. Finally, in either cases, the pinning length scales inversely
with the hydrophobic width $\tilde{w}_B$.

\begin{figure}[htb]
\centering\scalebox{0.33}{\includegraphics[clip=true]{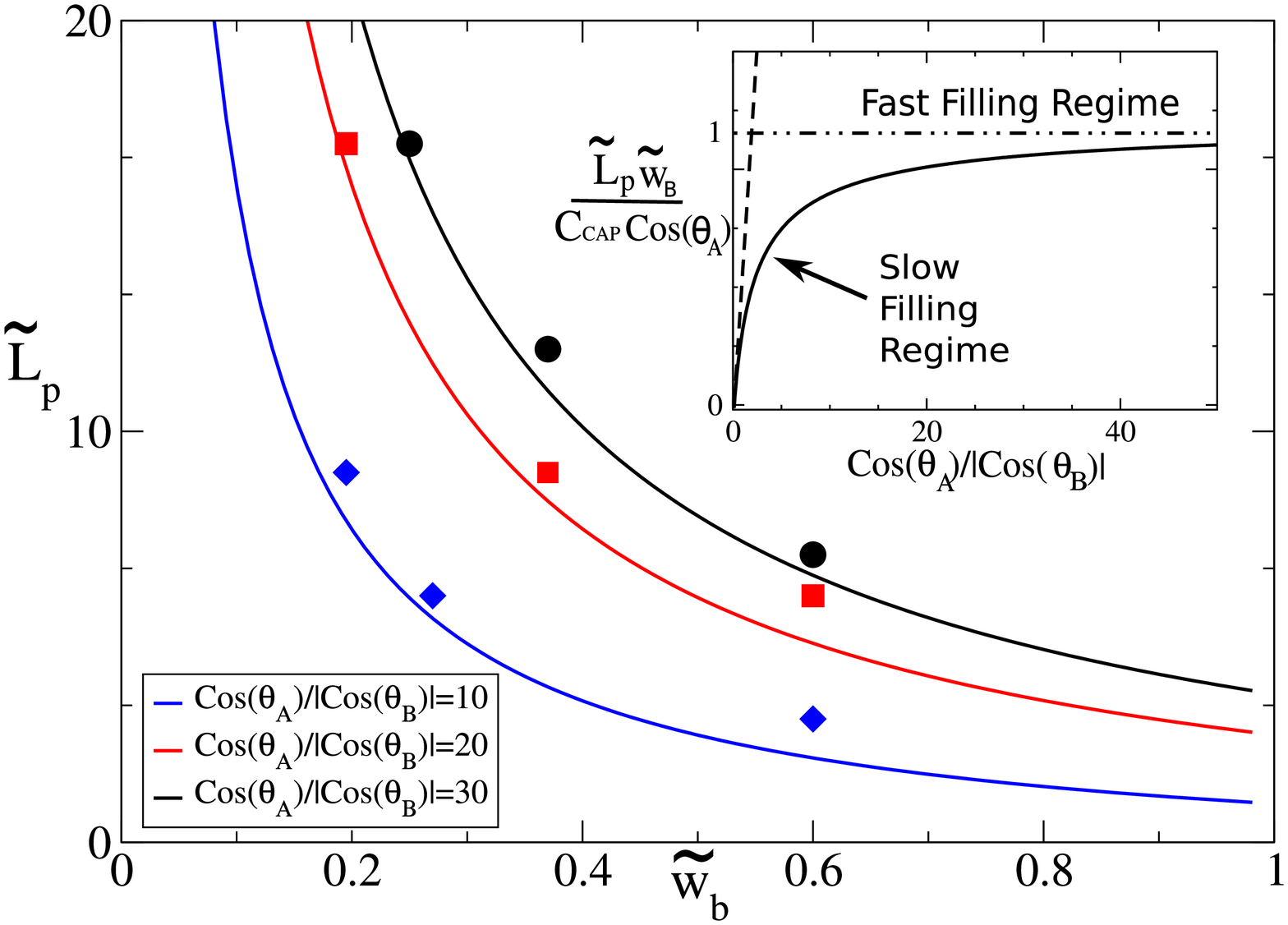}}
\caption{Normalized pinning length $\tilde{L}_p$ as a function of
  $\tilde{w}_B$ for $|\cos(\theta_A)/\cos(\theta_B)|=10,20,30$.  The
 continuous line represents the prediction given by
 Eq.~(\ref{eq:final_gen}), while the symbols are the result of the
 LB simulations. To match theory and simulation,
we have taken $\tilde{w}_B^{eff}=\tilde{w}_B-\tilde{\delta}$ in Eq.~(\ref{eq:final_gen}), where $\tilde{\delta}=\delta/H=0.025$. The other
  parameters used are: $\alpha=0.028, \tilde{L}=20, \tilde{C}_{cap}=5.7,\theta_B=92$.
 Inset: Normalized pinning length $\tilde{L}_p$ as a
function of $|\cos(\theta_A)/\cos(\theta_B)|$. Note the two distinct {\it fast filling} and
  {\it slow filling} regimes, as discussed in the text.}
\label{results_n}
\end{figure}

Eq.~(\ref{eq:final}) also shows that the quantity $\tilde{L}_p
\tilde{w}_B/ (C_{cap} \cos(\theta_A))$, is a universal function of
$\cos(\theta_A)/|\cos(\theta_B)|$ (see inset of Fig.\ref{results_n} ).
Since this function is bounded in the range [0:1],
Eq.~(\ref{eq:final}) clearly shows that $\tilde{L}_p$ is always
finite, unless $\tilde{w}_B=0$.

Let us notice that in presence of heterogeneous dilute coating ($w_A>>w_B$), one may use the
previous expression as a first order guess of the pinning length
$\tilde{L}_p$. For a typical case of a capillary filled with water at room
temperature, of width $H=10\mu m$, and with length of hydrophilic and
hydrophobic spots of the order of $\tilde{w}_B=0.01-0.1$, we obtain $\tilde{L}_p \simeq 50-500$.

\section{Discussion}
\begin{table}
 \begin{center}
\begin{tabular}{|| c | c |  c | c||}
\hline
$\delta /H$           & $n_{crit}^{th}$         &  $ n_{crit}^{sim}$    & $ n_{crit}^{th}/n_{crit}^{sim}$ \\ \hline
0.1           & 2               &7                   & 0.28 \\\hline

0.05           & 5                 & 7                  & 0.71 \\ \hline

0.033          & 8                    & 10                 & 0.83 \\ \hline

0.025           & 12                  &12                  & 1.\\ \hline

\end{tabular}
 \caption{Comparison between the prediction given by Eq.~(\ref{eq:final_gen}), based on the generalised Washburn equation~(\ref{washgendens}), and the
  LB simulation, expressed in terms of the number of hydrophobic
  wells $n_{crit}=[L_p/(w_A+w_B^{eff})-w_B^{eff}]$ that the front manages
  to overcome. The first column shows the ratio between the interface
  width $\delta$ and channel height $H$. The simulations were performed  taking into account the  correction due to the finite width of the interface, $w_B^{eff}=w_B-\delta$. The second
  column displays the number of wells $n_{crit}^{th}$ predicted by
  Eq.~(\ref{eq:final_gen}), while the third one refers to the LB
  results $n_{crit}^{sim}$. Theory and simulations converge towards
  the same prediction in the limit $\delta/H<0.03$.  }
\label{tab1}
\end{center}
\end{table}

The Lucas-Washburn equation (\ref{washgen}) does not
take into account two main sources of uncertainity. First, the inlet
dynamics may be sensitive to the structure of the reservoir,
leading to significant deviations from
the Poiseuille profile in the early stage of the filling process. Second, the description does not
take into account the dynamical effects induced on the interface by
the motion. On the other hand, it is
well known that a moving interface may be significantly distorted by the
viscous stress induced by the fluid motion, especially close to the contact
line. Such a dynamical bending, may require the introduction of a {\it
  dynamic contact angle} \cite{eggers,eggers2,cox,HuScriven71},
leading to a driving force in the Washburn law, which is itself
influenced by the interface motion, resulting in an unclosed "bootstrap" problem. Moreover, especially for hydrophilic coating, the shallow
fluid wedge at the boundary may induce an extra dissipation
\cite{degennes}, which modifies the dissipative terms. Asymptotically, the capillary speed becomes lower and lower,
and one may argue that, at large times, the assumptions behind
Eq.(\ref{washgen}) are increasingly well fulfilled, thereby justifying the neglect of
the above mentioned problems. Still, in many situations the asymptotic
``ideal'' regime is never reached, and consequentely one needs to assess the
influence of the previous effects on the filling process.

To check the validity of the "pinning criterion" on a realistic system, we
have resorted to a numerical simulations of a two-phase fluid, using Lattice
Boltzmann Equations\cite{pre1,pre2,Gladrow, Saurobook,bsv} in two
dimensions, to reproduce the capillary filling in presence of coated
patterns. The geometry is the same previously discussed
(Fig.~\ref{coat}), the only difference being the periodic boundary
conditions imposed at the two lateral sides in order to ensure total
conservation of mass inside the system. Similar problems, with or
without heterogeneous coating, have also been studied recently in
\cite{kwok,Latva07,Jia06, harting}.

  To consistently compare the simulation output with the
  theoretical prediction, we need to take into account the effect of
  two factors that differentiate simulations from the idealized description given by Eq.~(\ref{washgen}): (i) the presence
  of a finite liquid-gas density ratio $\alpha=\rho_g / \rho_l$; (ii) a
  finite width $\delta$ of the liquid-gas interface in the simulation.
  In order to take into account the
  unavoidable ``resistance'' of the gas occupying the capillary during
  the liquid invasion, one writes down the balance between the
  total momentum change inside the capillary and the force acting on
  the liquid$+$gas system. Simple calculations lead to a new equation
  governing the front dynamics \cite{condmat1,napoli}: 
\bea
  \label{washgendens} &[\alpha L + (1- \alpha) z] \frac{d^2 z}{d t^2}
  + (1-\alpha)(\frac{d z}{d t})^2 =\\ 
\nonumber & -\eta [\alpha L
  +(1-\alpha) z] \frac{d z}{d t}+f(z).  
\eea 
where $L$ is the total length of the channel and $\eta=12 \frac{\nu_l}{H^2(1+6
  \lambda_l/H)}$, being $\lambda_l$ the slip length of the fluid . In the limit $\alpha\rightarrow 0$ and $\lambda_l \rightarrow 0$,
one recovers the original Eq.~(\ref{washgen}). Equation~(\ref{washgendens}) can be mapped onto the
same Eq.~(\ref{basic}), only with a slightly different definition of
$q=[\alpha L +(1-\alpha) z]^2/2$,  and $F(q)=2 (1-\alpha) V_{cap} V_d \cos(\theta[z(q)])$.

 The pinning criterion Eq.~(\ref{eq:final})
therefore simply changes to
\begin{equation}
\label{eq:final_gen}
\tilde{L}_p\simeq \frac{\tilde{C}_{cap}\; cos(\theta_A)}{\tilde{w}_B}\left[1-\frac{ \log \left(1+\frac{\cos(\theta_A)}{|\cos(\theta_B)|}\right)}{\frac{cos(\theta_A)}{|cos(\theta_B)|}} \right]-\frac{\alpha}{(1-\alpha)} \tilde{L}.
\end{equation}   
where $\tilde{C}_{cap}=C_{cap}(1+6 \lambda_l/H)^2/(1-\alpha)$ and $\tilde{L}$ is the dimensionless channel length $\tilde{L}=L/H$. This generalised criterion, denoted from now on as $\tilde{L}_p^{th}$, is tested against LB
  simulations.

The above equation is exact if evaporation-condensation effects are
negligible, i.e. when the gas is pushed out of the capillary, without
any interaction with the liquid. This is not the case for most
mesoscopic models available in the literature \cite{shanchen,yeomans},
which are based on a diffusive interface dynamics
\cite{jacqmin,Seppecher96,pismen,pismen2}.  As shown in a previous paper
\cite{condmat1}, the dynamics given by (\ref{washgendens}) is
correctly recovered only in the limit of thin interface $\delta/H
\rightarrow 0$ and of negligible gas-liquid density ratio, $\alpha
\rightarrow 0$

For the present case, the existence of a finite interface width
$\delta$ (generally around 5-6 grid points) in the simulation, has the
effect of introducing an effective length of the hydrophobic obstacle
$w_B$. Therefore, in the following, we compare the simulation results
with the solution of Eq.~(\ref{washgendens}), with an effective
length of the obstacle $w_B^{eff}=w_B-\delta$. Besides, in order to
recover the correct hydrodynamic limit, the additional requirement
$\delta/H<<1$ must be fullfilled (Table 1).

\begin{figure}[htb]
\centering\scalebox{0.35}{\includegraphics[clip=true]{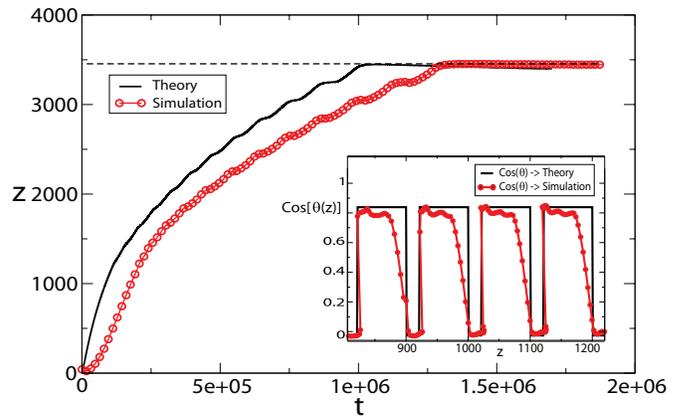}}
\caption{Time evolution of the front coordinate in the LB simulation (red
  points) and as numerical solution of Eq.~(\ref{washgendens}) (black
  line).  The simulations were performed using the following
  parameters: $\alpha=0.028, L=3600, H=200, w_A=144, w_B=56,
  \gamma=0.14, \nu= 1/6$. For this case $\delta/H=0.025$, in lattice
  units. The dashed line marks the prediction of Eq.~(\ref{eq:final_gen}). Inset: comparison between imposed
  (black line) and observed (red dots) contact angles: note that
  while the imposed angle $\cos(\theta (z))$ is a step-wise
  function, the simulation clearly shows an hysteresis
  \cite{hysteresis1,hysteresis2} of the contact angle. The observed
  contact angle was obtained by interpolating the front shape with a
  circle of radius $R$, so that $\cos(\theta)=H/2R$. Note that, due to
  the convex shape of the interface, the front coordinate (calculated
  at $y=H/2$) always precedes the location where the front first meets
  the hydrophobic region.  }
\label{comparison}
\end{figure}

\section{Simulation Results}
 Fig.~\ref{results_n} shows the
prediction for the dimensionless
 pinning length $\tilde{L}_p^{th}$
as a function of the two main
 parameters $\tilde{w}_B$ and
$\cos(\theta_A)/|\cos(\theta_B)|$
. The main figure displays
three sets of results, corresponding
 respectively to
$\cos(\theta_A)/|\cos(\theta_B)|=10,20,30$. The solid lines
represent the numerical solutions of Eq.~(\ref{eq:final_gen}), while the symbols correspond to
the LB simulation results. As one can see, the numerical results
 are
in satisfactory agreement with the theoretical
 prediction. To be
noted that we have limited the numerical comparison up to pinning
lenghts $\tilde{L}_p^{max} \simeq 15-20$ because of computational
constraints.  

In the inset, the universal curve $\left[ 1-  \frac{\log \left(1+
\frac{\cos(\theta_A)}{|\cos(\theta_B)|}\right)}{\frac{cos(\theta_A)}{|cos(\theta_B)|}} \right]$ is reported:
it is possible to appreciate the two different regimes of {\it slow
  filling}, in which the product $\tilde{L}_p\tilde{w}_B/ (C_{cap} \cos(\theta_A))$ grows linearly with
$\cos(\theta_A)/|\cos(\theta_B)|$, and {\it fast filling}, in which this quantity saturates to 1.

It must be noted that the convergence towards $\tilde{L}_p^{th}$ is
accomplished only when the width $\delta$ of the liquid-gas interface
becomes thin enough with respect to the characteristic length $H$ of
the system (Table 1). In this regime, corresponding to
$\delta/H<0.03$, the overall asymptotic trend and the final
theoretical pinning position are in agreement with the
simulation, as shown in Fig.~\ref{comparison}.  The discrepancy
between the two curves during the initial transient regime is due to the ``vena contracta'' phenomenon, reflecting the non-trivial
matching between the reservoir and the capillary dynamics at the inlet
of the simulation setup \cite{venacontracta}. Such small discrepancy
can also be reabsorbed into an added mass term in
LHS of Eq.(\ref{washgen}), as shown quantitatively in \cite{condmat1}.

At larger values of the ratio $\delta/H$, the error on $\tilde{L}_p$
becomes higher, and in these cases the simulation systematically
gives $\tilde{L}_p > \tilde{L}_p^{th}$ (Table~\ref{tab1}). Such
discrepancy is connected with the departure from hydrodynamics,
for length scales of the order of the interface width, typical of any
diffuse-interface description. 

The satisfactory agreement between the LB simulations and the hydrodynamical description
based on the generalized Washburn equation (\ref{washgendens})
indicates that effects induced by the distortion of the interface
profile are negigible, at least for the choice of coating
properties discussed in this work.  On the basis of simple hydrodynamical considerations, one
would expect that the extra dissipation term induced by the wedge
close to the contact line should introduce a correction
\cite{voinov,cox,eggers}:
\begin{equation}
F_{ex} \propto \frac{ \dot z}{tg(\theta)} \log(H/\lambda)
\label{eq:extra}
\end{equation}
where $\lambda$ is the ultraviolet cutoff (inner length scale) up to
which hydrodynamics applies. In LB simulations, it is reasonable to assume $\lambda \sim \delta$. The
correction term (\ref{eq:extra}), being localized at the interface, is always sub-leading at long times,
 differently from the viscous
dissipation that is proportional to the volume occupied by the
fluid. This extra-stress can  change the local interface
profile, in particular close to the transition between two regions
with different coating properties. This is shown in the inset of
Fig.~\ref{comparison}, where the  effective {\it dynamic} contact angle,
 is plotted versus the static (microscopic)
contact angle imposed by the boundary conditions. As seen, the dynamic
profile is departing from the static one only when the front enters the
hydrophobic region, i.e.  where the interface tends to be pinned.
Eq.~(\ref{eq:extra}) suggests that  extra dissipation has an effect
also for long times, in the case of filling with very hydrophilic walls ($\theta_A
\sim 0^o$), and/or of filling in viscoelastic flows, depending on the
rheological properties of the fluid. These interesting issues are left for future research.

\section{Conclusions}
In conclusion, we have derived an analytical criterion
(Eq.~(\ref{eq:final_gen})) to predict the pinning location
  of a moving front in a 2d plane geometry, based on the generalized
  Lucas-Washburn equation. This criterion has been derived under the assumption of diluteness of hydrophobic spots, and has been
  satisfactorly tested against 2d Lattice Boltzmann simulations. We
  have also discussed limitations of both approaches: the first due to the neglect of interface distortion and extra-dissipation at the
  contact line for Washburn-like equations, and the second 
  regarding the effects induced by deviation from hydrodynamics in LBE
  models. We have shown that, in the limit of fast filling process
  ($\theta_A >>0$), and for sufficiently thin interfaces, the
  analytical criterion and the LB simulations are in satisfactory
  agreement with each other.

Work to extend the pinning criterion to more complex situations, such as in
the presence of stochastic heterogeneous coating (not diluted) and/or
in the presence of geometrically coated interfaces, is currently
underway.

\acknowledgments Financial support through the EC project INFLUS is
kindly acknowledged. Valuable discussions with M. Sbragaglia, F. Toschi and J. Yeomans are kindly
acknowledged.

\end{document}